\documentclass[tighten, times, twocolumn]{aastex631}

\usepackage{amsmath}
\usepackage{amsfonts} 
\usepackage{graphicx,color}

\newcommand\method{RAM}

\begin{document}

\title{{\method}: Rapid Advection Algorithm on Arbitrary Meshes}

\correspondingauthor{Pablo Ben\'itez-Llambay}
\email{pablo.benitez@uai.cl}

\author[0000-0002-3728-3329]{Pablo Ben\'itez-Llambay}
\affiliation{Facultad de Ingenier\'ia y Ciencias, Universidad Adolfo Ib\'añez, Av. Diagonal las Torres 2640, Peñalol\'en, Chile}
\affiliation{Data Observatory Foundation, ANID Technology Center No. DO210001}
\author[0000-0001-7671-9992]{Leonardo Krapp}
\affiliation{Department of Astronomy and Steward Observatory, University of Arizona, Tucson, AZ 85721, USA}
\author[0000-0002-2450-7389]{Ximena S. Ramos}
\affiliation{Millennium Institute of Astrophysics MAS, Nuncio Monseñor Sotero Sanz 100, Of. 104, Providencia, Santiago,
Chile}
\author[0000-0001-5253-1338]{Kaitlin M. Kratter}
\affiliation{Department of Astronomy and Steward Observatory, University of Arizona, Tucson, AZ 85721, USA}

\begin{abstract}
The study of many astrophysical flows requires computational algorithms that can capture high Mach number flows, while resolving a large dynamic range in spatial and density scales. In this paper we present a novel method, RAM: Rapid Advection Algorithm on Arbitrary Meshes. RAM is a time-explicit method to solve the advection equation in problems with large bulk velocity on arbitrary computational grids.
In comparison with standard up-wind algorithms, RAM enables advection with larger time steps and lower truncation errors.
Our method is based on the operator splitting technique and conservative interpolation. 
Depending on the bulk velocity and resolution, RAM can decrease the numerical cost of hydrodynamics by more than one order of magnitude.
To quantify the truncation errors and speed-up with RAM, we perform one and two-dimensional hydrodynamics tests.
We find that the order of our method is given by the order of the conservative interpolation and that the effective speed up is in agreement with the relative increment in time step.
RAM will be especially useful for numerical studies of disk-satellite interaction, characterized by high bulk orbital velocities, and non-trivial geometries. Our method dramatically lowers the computational cost of simulations that simultaneously resolve the global disk and well inside the Hill radius of the secondary companion.
\end{abstract}

\section{Introduction}
\label{sec:introduction}
A broad range of open questions in astrophysics rely on the non-linear outcome of computational studies of fluid dynamics. These problems often involve high Reynolds number flows that must be studied across many orders of magnitude in scale.
A central aspect of computational fluids dynamics is finding stable, accurate, conservative and efficient numerical methods to solve the advection equation
\begin{equation}
\partial_t Q + \partial_x \left( Q v\right) = 0\,,
\label{eq:advection}
\end{equation}
for a given initial condition $Q(x,t=0) = Q_0(x)$ and suitable boundary conditions. Here, $Q(x,t)$ is the advected quantity along the $x$ direction by the velocity $v$ over a time $t$.
Without loss of generality, we can assume $v=v_0+\delta{v}$, where the bulk velocity, $v_0$, is constant in space and $\delta v(x)$ is the departure of the flow velocity $v$ from $v_0$.

In the Eulerian formalism, Eq.\,\eqref{eq:advection} can be solved through the utilization of Riemann or finite volume algorithms combined with upwind reconstruction methods, like piece-wise linear \citep[e.g.][]{VanLeer1977} or piece-wise parabolic \citep{Colella1984} methods, among others. 
These conservative upwind methods introduce truncation errors that depend on the bulk velocity.
Such methods introduce a direct correlation between the accuracy of the numerical solution and the bulk velocity of the flow \citep[see e.g.,][]{Robertson2010}.
Therefore, solving the advection equation in a frame of reference that moves at a speed equal to the bulk velocity minimizes the errors in the solution.

In addition, explicit finite differences methods give rise to the so-called Courant–Friedrichs–Lewy (CFL) stability condition, which is inversely proportional to the velocity $v$. As such, a large bulk velocity limits the allowed time step $\Delta t$, therefore increasing the numerical cost of finding the solution at time $t$. 
Crucially, this numerical cost is unbounded as it scales with the - arbitrarily large - bulk velocity $v_0$. In contrast, finite volume methods do not require a CFL condition as long as the upwind fluxes are integrated over the entire domain of dependence. However, it is customary to limit the distance that information can travel within a time-step such that it is less than the size of the smaller cell.

Because of these two observations, dealing with a large bulk speed requires the development of special techniques first, to decrease the large truncation errors and, second, to increase the effective time step.
Increasing the time step (without compromising the stability of the solver) is an especially effective way to decrease the numerical cost of the simulations.

In terms of $v_0$, Eq.\,\eqref{eq:advection} can be written as
\begin{equation}
\partial_t Q + v_0 \partial_x  Q + \partial_x \left( Q \delta v\right) = 0\,.
\label{eq:advection_bulk}
\end{equation}
In order to address the problems described above, it is desired to remove $v_0$ from Eq.\,\eqref{eq:advection_bulk}.
One way of doing that is to define the coordinate transformation 
\begin{equation}
\begin{aligned}
t'&=t\,, \\
x'&=x-v_0 t\,,
\end{aligned}
\end{equation}
such that Eq.\,\eqref{eq:advection_bulk} becomes
\begin{equation}
\partial_{t'} Q + \partial_{x'} \left( Q \delta v\right) = 0\,.
\label{eq:advection_moving}
\end{equation}
This approach has been followed by \cite{Shariff2018} to solve advection on fixed uniform meshes. It consists in solving advection via standard methods in the frame of reference that moves at the bulk velocity and, at the end of each time step, return to rest frame and fill a fixed mesh by shifting the advected quantity an amount $v_0\Delta t$ in Fourier space \citep[see also][]{Johansen2009}. This can be done efficiently through the utilization of the Fast Fourier Transform (FFT) method. 
However, an FFT (and its inverse) can only be done efficiently on meshes that are uniform.
In non-uniform spaces, a Non-Uniform Discrete Fourier Transform (NUDFT) can be used, but at the expense of additional cost or precision \citep{KKP2009}.
Moreover, arbitrary shifting of non-smooth functions through the Discrete Fourier Transform is subject to the Gibbs phenomenon, which produces over/undershoots in the reconstructed (shifted) signal near large gradients in the sampled function.
More importantly, Gibbs phenomenon produces spurious values that could change sign of the advected quantity. This could be particularly catastrophic for positive-valued quantities like the density or internal energy.

A second possibility to remove $v_0$ from Eq.\,\eqref{eq:advection_bulk}, also related to a coordinate transformation, is to solve the advection equation in a Lagrangian mesh that moves at the bulk speed, as done for example by the codes AREPO \citep[][]{Springel2010} or DISCO \citep{Duffell2016}. In one dimension this is, perhaps, the optimal solution on both uniform and non-uniform domains. However, the time step could be limited in multidimensional meshes with strong shear unless the mesh is regularly connected.

Another possibility is the FARGO algorithm developed by \citep{Masset2000}\footnote{FARGO, or Fast Advection for Rotating Gaseous Objects, is an advection algorithm that shares its name with the original hydrodynamics code in which it was implemented. Here, when we refer to FARGO, we mean the algorithm, rather than the namesake code.}.
The FARGO algorithm is a widely used approach in the context of disk dynamics to avoid the limitations of the time step due to a large bulk speed in differentially rotating fluids.
The FARGO algorithm employs the operator splitting method on a static mesh (or one rotating as a rigid body) \citep[see e.g.,][]{Masset2000, BenitezLlambay2016}, for which, to first order in time, Eq.\,\eqref{eq:advection_bulk} is solved in two separate steps.
In a first step, one solves
\begin{equation}
    \partial_t Q + v_0 \partial_x Q = 0\,,
    \label{eq:splitting_first}
\end{equation}
and then, using the solution of Eq.\,\eqref{eq:splitting_first} as an initial condition, one solves through standard methods
\begin{equation}
    \partial_t Q + \partial_x \left( Q \delta v\right) = 0\,.
    \label{eq:splitting_second}
\end{equation}
Solution of Eq.\,\eqref{eq:splitting_second} thus corresponds to the approximate solution of Eq.\,\eqref{eq:advection_bulk}. To first order, this method is equivalent to the one proposed by \citet{Shariff2018}. One crucial property of the FARGO algorithm is that Eq.\,\eqref{eq:splitting_first} has an analytical solution
\begin{equation}
Q(x,t)= Q(x-v_0 t, 0)\,,
\label{eq:advection_analytic}
\end{equation}
which corresponds to a shift of the initial condition along the $x-$coordinate. On uniform meshes, this shift can be cast as an integer shift of grid cells plus a residual advection. Furthermore, on periodic domains the integer shift corresponds to a cyclic permutation of grid cells, so it does not introduce any numerical error. 
While the simplicity of this integer shift makes the FARGO algorithm extremely efficient, it is not applicable to non-uniform meshes.

\medskip

In this work, we present an algorithm to solve Eq.\,\eqref{eq:advection} with arbitrarily large bulk speed $v_0$ that brings the efficiency of the FARGO algorithm to uniform and non-uniform meshes alike.
Our method is based, as FARGO, on the operator splitting technique and the method of characteristics with suitable conservative reconstruction methods that are used to evaluate Eq.\,\eqref{eq:advection_analytic}.
Also, the method of characteristics that we develop to shift the signal could be used to build higher-order schemes on non-uniform meshes (or to avoid potential issues with FFT methods on uniform meshes).
We  furthermore show that if the underlying mesh is distributed according to a predefined mesh density function, a coordinate transformation that maps the non-uniform space into a uniform one speeds up calculations significantly.

\bigskip 

This paper is organized as follows. In Section \ref{sec:definitions} we define the basic equations used throughout the manuscript. In Section \ref{sec:method} we present the details of the RAM method for solving the advection equation. 
Results from test problems and benchmarking of our implementation are shown in Section \ref{sec:tests}. Finally, we summarize our work in Section \ref{sec:conclusions}.

\section{Basic Definitions}
\label{sec:definitions}

Consider a static grid made of $N+1$ nodes with monotonic coordinates $x_{\rm min} = x_{-\frac{1}{2}} < x_{\frac{1}{2}} < \dots < x_{N-\frac{1}{2}} = x_{\rm max}$. We assume the function $Q$ to be periodic within the domain and sampled at the cell coordinates $x_i = (x_{i+\frac{1}{2}}+x_{i-\frac{1}{2}})/2$ at time $t_n$ and define $Q^n_i \equiv Q(x_i, t_n)$. Consecutive pairs of nodes define the cell size $\Delta x_i = x_{i+\frac{1}{2}}-x_{i-\frac{1}{2}}$ for $i=0, \dots, N-1$. The volume-average of $Q$ within a cell is
\begin{equation}
 \langle Q \rangle_i^n  \equiv \frac{1}{\Delta x_i} \int_{x_{i-\frac{1}{2}}}^{x_{i+\frac{1}{2}}} Q(x,t_n)\, {\rm d}x\,.
 \label{eq:cellaverage}
\end{equation}
In this work, we assume 
\begin{equation}
\langle Q \rangle_i^n = Q_i^n\,.
\label{eq:cellaverage_assumption}
\end{equation}
It can be shown via a Taylor expansion of $Q$ around $x_i$ that, for smooth functions, Eq.\,\eqref{eq:cellaverage_assumption} corresponds to a second-order approximation of Eq.\,\eqref{eq:cellaverage} within the $i$-cell.
Therefore, the volume integral of $Q$ within a cell is given by
\begin{equation}
M^n_i \equiv Q^n_i \Delta x_i\,.
\label{eq:Qmass_i}
\end{equation}
For a periodic domain, a conservative update of $Q$ from time $t=t_0$ to $t=t_n$ implies
\begin{equation}
\sum_{i=0}^{N-1} M^n_i  = \sum_{i=0}^{N-1} M^{n-1}_i = \cdots = \sum_{i=0}^{N-1} M^0_i\,.  
\label{eq:qmass_total}
\end{equation}

Because $Q$ is sampled at a discrete set of points, we also define
\begin{equation}
q^n(x) \equiv  q_i^n(x) \quad {\rm for} \quad x_{i-\frac{1}{2}} \leq x < x_{i+\frac{1}{2}}\,.
\end{equation}
The function $q^n$ is made of a set of piece-wise interpolating monotonic functions $q_i(x,t_n)$ that approximate $Q(x,t_n)$ within each cell, such that
\begin{equation}
    \sum^{N-1}_{i=0} \int_{x_{i-\frac{1}{2}}}^{x_{i+\frac{1}{2}}} q_i^n(x) \,{\rm d} x = \sum^{N-1}_{i=0} M_i^n \,.
     \label{eq:Qmass_interp}
\end{equation}
In particular, we also assume
\begin{equation}
     M_i^n = \int_{x_{i-\frac{1}{2}}}^{x_{i+\frac{1}{2}}} q_i^n(x) \,{\rm d} x \,.
     \label{eq:Qmass_interp_1}
\end{equation}
Furthermore, it is required that $q$ does not introduce new extrema, preventing the development of over/under shoots in the solution. 

In terms of $q$, Eq.\,\eqref{eq:advection_analytic} becomes
\begin{equation}
    q(x,t) = q(x-v_0 t, 0)\,.
    \label{eq:advection_analytic_q}
\end{equation}
With the above relation, the advection of the discrete function $Q$ is transformed into the advection of a piece-wise, continuous function $q$ that shares the local and global volume integral of $Q$.

Finally, given a velocity $v(x)$, it is customary to define the constant background bulk velocity $v_0$ as the spatially averaged velocity \citep[c.f.][]{Masset2000}. This definition is arbitrary; for example, it could be defined as the velocity of the center of mass. To obtain an optimal solution both in terms of efficiency and accuracy, $v_0$ should be defined such that it minimizes the maximum amplitude of $|\delta v| = |v-v_0|$ within the domain, so the allowed time step is maximized. This is readily obtained after defining $v_0$ as the average between the maximum and minimum of $v$ \citep[see][section 3.5.1]{BenitezLlambay2016}. 
It is worth noticing that the method presented in this paper is agnostic about the definition of $v_0$ as long as it is constant in space. The way it is defined influences the effective speed up and overall precision obtained.

\section{{\method}}
\label{sec:method}
As discussed previously in Section \ref{sec:introduction}, a method to solve Eq.\,\eqref{eq:advection} that maximizes the time step and minimizes truncation errors in arbitrary meshes requires an efficient solution of Eq.\,\eqref{eq:splitting_first}. In other words, we need a method to evaluate the exact solution given by Eq.\,\eqref{eq:advection_analytic} that is fast, stable and conservative. 

Evaluating Eq.\,\eqref{eq:advection_analytic} from time $t_n$ to $t_{n+1}=t_n+\Delta t$ corresponds to shifting $Q(x,t_n)$ an amount $v_0 \Delta t$ in space.
In the particular case where $Q$ is sampled on a discrete but uniformly spaced grid and $v_0 \Delta t$ is an exact multiple of the cell size, the shift is achieved via a cyclic permutation of indices \citep[see][]{Masset2000}. However, in general, $Q$ could be sampled on a non-uniform domain. In this case, it is impossible that for every cell, the cell size is an integer multiple of $v_0 \Delta t$. Therefore, shifting $Q$ requires the utilization of interpolation methods. 
Here we provide a method to evaluate Eq.\,\eqref{eq:advection_analytic} via Eq.\,\eqref{eq:advection_analytic_q} that is conservative, i.e., it conserves the total volume average of $Q$ to machine precision (as defined by Eq. \ref{eq:qmass_total}).

\medskip

\begin{figure}[t]
\centering
\includegraphics{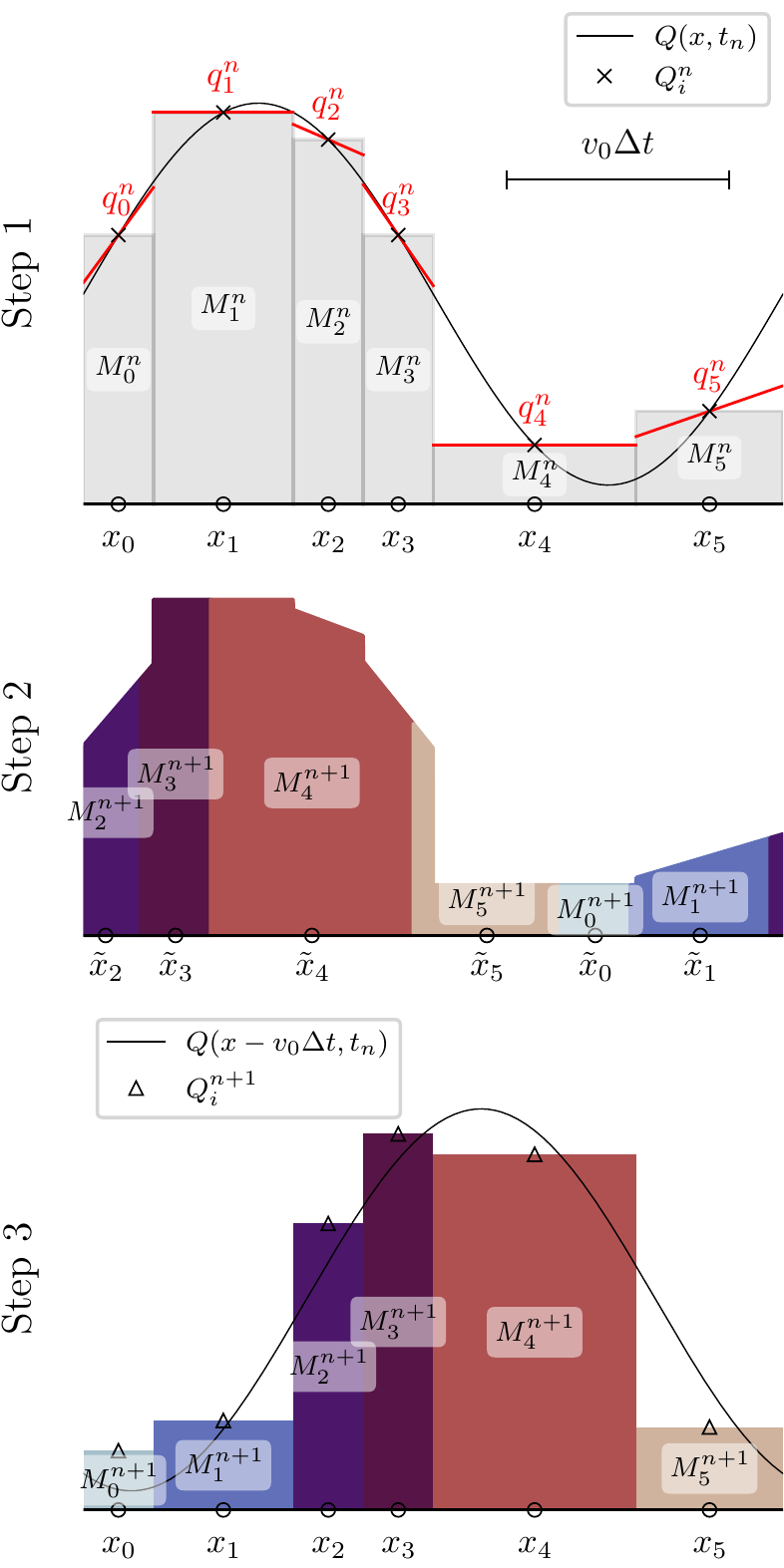}
\caption{{\method} steps involved in solving Eq.\,\eqref{eq:splitting_first} for a positive valued velocity $v_0$ over a time $\Delta t$ in a periodic domain. The upper panel shows an arbitrary continuous function $Q(x)$ at time $t_n$, known at a discrete set of points $x_i$, with value $Q_i^n$. These values allow us to calculate $q_i^n$, which we draw with red lines. For this particular example, we use PLM interpolation (see Section\,\ref{sec:step1}). Here, we also draw the length $v_0 \Delta t$ used to shift the mesh to the left in step 2.
The middle panel shows $M_i^{n+1}$ obtained from evaluating Eq.\,\eqref{eq:mass_shift_algo}, where different colors are used for different cells. These colors allow us to link each area (integral within each cell of $q^n$) with the resulting area obtained when reconstructing $Q_i^{n+1}$ in step 3 (see Section\,\ref{sec:step2}).
The lower panel shows the reconstructed values $Q_i^{n+1}$ obtained from $M_i^{n+1}$ via Eq.\,\eqref{eq:Qmass_i}. We also show the exact solution $Q(x-v\Delta t, t_0)$ for completeness.}
\label{fig:method}
\end{figure}

To evaluate Eq.\,\eqref{eq:advection_analytic_q} from time $t_{n}$ to time $t_{n+1}$, we propose the following 3 steps
\begin{enumerate}
    \item Calculate the piece-wise interpolating functions $q_i(x, t_n)$ using the discrete nodes $Q_i^n$ (see Section \ref{sec:step1}).
    \item Obtain $M_i^{n+1}$ after shifting analytically $q^n$ an amount $v_0\Delta t$ (see Section \ref{sec:step2}).
    \item Use Eq.\,\eqref{eq:Qmass_i} to obtain the advected profile (see Section \ref{sec:step3}).
\end{enumerate}
\medskip
These three steps are shown explicitly in Figure \ref{fig:method}. Together with the operator splitting needed to build Eq.\,\eqref{eq:splitting_first}, these three steps are the core of our Rapid Advection Algorithm on Arbitrary Meshes ({\method}). In what follows, we explain each of these step and give a practical example.

\subsection{Step 1 - Interpolating $Q_i^n$}
\label{sec:step1}
The first step of the {\method} method requires the interpolation of the nodes $Q^n_i$ to obtain the  piece-wise functions $q_i^n(x)$. 
Ideally, this interpolation should be of the order (or higher) than the one used to solve Eq.\,\eqref{eq:splitting_second}, so errors are dominated by the standard advection step.
The {\method} method does not rely on any specific choice as long as the interpolation is conservative and does not introduce new extrema.
In this work we explore two different conservative piece-wise monotonic interpolation methods (i.e., do not introduce new maxima/minima) that are often utilized in hydrodynamics. These are the Piece-wise Linear (PLM) and Piece-wise Parabolic (PPM) Methods that we explain below. Throughout this section, all the quantities are evaluated at time $t_n$, so for the sake of simplicity, we omit the superscript $n$.

\paragraph{Piece-wise Linear Method}
This method provides an interpolation that is accurate up to second order  \citep[see e.g,][section 3.1]{Mignone2014}. Assume $q_i$ to be of the form
\begin{equation}
    q_i(x) = Q_i + \sigma_i \left( x - x_i \right)\,,
\end{equation}
where $\sigma_i$ is the slope of the linear function. Monotonicity is preserved by utilizing the slope limiter proposed by \cite{VANLEER1974}, where

\begin{equation}
   \sigma_i = 
  \begin{cases}
       \displaystyle{0} \quad\quad\quad &\text{if }  \Delta Q_{i+\frac{1}{2}}\Delta Q_{i-\frac{1}{2}} \leq 0\\
       \\
       \displaystyle{\frac{2 \Delta Q_{i+\frac{1}{2}}\Delta Q_{i-\frac{1}{2}}}{ \Delta Q_{i+\frac{1}{2}} + \Delta Q_{i-\frac{1}{2}} }}  \quad &\text{otherwise } 
     \end{cases}
\end{equation}
with $\Delta Q_{i+\frac{1}{2}} = (Q_{i+1}-Q_i) / (x_{i+1} - x_i)$.

\medskip

\paragraph{\bf Piece-wise Parabolic Method}

This method, first introduced by \cite{Colella1984}, interpolates $Q$ up to third-order via quadratic polynomials. The piece-wise parabolic function is
\begin{align}
    q_i(x) &=  Q_{L,i} + \left( \frac{x-x_i}{\Delta x_i} \right)\left( Q_{R,i} - Q_{L,i} \right) \nonumber \\
    &+  Q_{6,i} \left(1-\frac{x-x_i}{\Delta x_i}\right) \left( \frac{x-x_i}{\Delta x_i} \right)\,
\end{align}
with $Q_{6,i} =  6\langle Q \rangle_i - 3\left( Q_{L,i} + Q_{R,i} \right) $. 
To obtain the quadratic polynomials we need to
calculate $\langle Q \rangle_i$ (see Section\,\ref{sec:definitions}) and interpolate $\langle Q \rangle_i$ to the cell interfaces to obtain $\langle Q\rangle_{i+\frac{1}{2}} =  Q_{{\rm L},i+1} = Q_{{\rm R},i}$ (see Appendix\,\ref{ap:PPM}). Furthermore, monotonicity constraints have to be applied (see Appendix\,\ref{ap:monotonocity}).
It is worth noticing that the coefficients of the quadratic polynomial depend on $\langle Q \rangle_{i}$, $Q_{{\rm L},i}$, and $Q_{{\rm R},i}$, therefore the errors of the latter limit the accuracy of the interpolation.
In this work we assume $\langle Q\rangle_i \equiv Q_i$ which is a second-order approximation of $\langle Q \rangle_i$.
Therefore, we expect our interpolation errors to be smaller than third-order.
Higher accuracy could be obtained by improving the calculation of $\langle Q \rangle_i $ \citep[][]{Colella2008}. 
Note, however, that since PPM conserves $\langle Q \rangle_i $ locally, a higher order approximation of $\langle Q \rangle_i $ may not be conservative in the terms defined in this work, i.e., it does not conserve $\sum_i Q_i \Delta x_i$ over successive time steps. The conservation will depend upon the  approximation of the  spatial derivative of $Q$. For example, a fourth-order approximation of $\langle Q \rangle_i  =  Q_i + 1/24 \,\, (Q_{i+1} - 2Q_i + Q_{i-1})  $ is conservative globally over a periodic domain (i.e., it satisfies Eq.\,\ref{eq:qmass_total}).
In general, the accuracy of our method can be increased provided a higher order approximation of $\langle Q \rangle_i $ is conserved within a time step \citep[e.g,][]{Felker2018}.

\medskip

Finally, we note that for both PLM and PPM, monotonicity constraints increase truncation errors locally within non-smooth regions which, in turn, reduces the overall accuracy of the interpolation. An example of reduction of accuracy due to this effect is shown in Section\,\ref{sec:1d-advection}, for the advection of a square profile.

\subsection{Step 2 - Obtaining $M^{n+1}$ from the advection of $q^n$}
\label{sec:step2}

The volume integral $M_i^{n+1}$ is obtained from the definition \eqref{eq:Qmass_i} combined with the solution \eqref{eq:advection_analytic_q}
\begin{align}
    M_i^{n+1} &= \int^{x_{i+\frac{1}{2}}}_{x_{i-\frac{1}{2}}} q(x- v_0\Delta t, t_n) {\rm d}x\,.
    \label{eq:Mnp}
\end{align}
After defining 
$\tilde{x} \equiv x - v_0\Delta t$,
we can write Eq.\,\eqref{eq:Mnp} as
\begin{align}
    M_i^{n+1} =
    \int^{\tilde{x}_{i+\frac{1}{2}}}_{\tilde{x}_{i-\frac{1}{2}}} q^n(x) {\rm d}x\,.
    \label{eq:mass_advected}
\end{align}
We define positive integers, $p(i)$, such that
\begin{equation}
    x_{p(i)-\frac{1}{2}} \leq \tilde{x}_{i-\frac{1}{2}}  < x_{p(i)+\frac{1}{2}}\,,
    \label{eq:condition_p}
\end{equation}
which allows Eq.\,\eqref{eq:mass_advected} to be evaluated as
\begin{equation}
 M^{n+1}_i = 
     \begin{cases}
       \displaystyle{\int^{\tilde{x}_{i+\frac{1}{2}}}_{\tilde{x}_{i-\frac{1}{2}}} q^n_{p(i)} {\rm d}x}, \quad \text{if} \quad p(i) = p(i+1) \\
       \displaystyle{\int^{x_{p(i)+\frac{1}{2}}}_{\tilde{x}_{i-\frac{1}{2}}} q^n_{p(i)} {\rm d}x + \sum^{m<p(i+1)}_{m=p(i) +1} M^{n}_m}   \\       +\displaystyle{\int^{\tilde{x}_{i+\frac{1}{2}}}_{x_{p(i+1)-\frac{1}{2}}} q^n_{p(i+1)} {\rm d}x},  \quad \text{otherwise } 
     \end{cases}
     \label{eq:mass_shift_algo}
\end{equation}
where $p(N) = p(0)$ due to periodicity. Here, the sum over $M_m^n$ corresponds to the integral of $q_m$ over an entire cell centered at $x_m$. Because the reconstruction is conservative, we replace this integral with $M_m^n$. 

Eq.\,\eqref{eq:mass_shift_algo}  is the generalization of standard upwind methods for arbitrary domains of dependence and can be interpreted as a conservative remapping of volume-averaged values (e.g., mass) from coordinates $x_i$ to the new coordinates $\tilde{x}_i = x_i - v_0\Delta t$. Note that since $\Delta t$ is unbound, $\tilde{x}_i$ may be nowhere near $x_i$, and/or a grid cell in $x$ may intersect one, two, or multiple grid cells in $\tilde{x}$ and vice-versa.  
 However, this remapping is simplified by the index calculation $p(i)$ (Eq.\,\eqref{eq:condition_p} and Section\,\ref{sec:mesh-density}) and the two conditionals in Eq.\,\eqref{eq:mass_shift_algo}.
If the time step $\Delta t$ is bounded (e.g., $\Delta t \lesssim {\rm min}(\Delta x_i)/v_0$), Eq.\,\eqref{eq:mass_shift_algo} reduces to the coordinate remap described in \citet{Lufkin1993} (see their equation 58). In this case, $p(i)=i$, $p(i+1)=i+1$, and the algorithm in Eq.\,\eqref{eq:mass_shift_algo} can be cast into the form of a standard upwind advection scheme, similar to the one used to solve Eq.\,\eqref{eq:splitting_second}.

\subsubsection{Efficient calculation of $p(i)$}
\label{sec:mesh-density}

If it happens that a coordinate transformation $u(x)$ exists such that $\Delta u \equiv u(x_{i+\frac{1}{2}})-u(x_{i-\frac{1}{2}})$ is constant, Eq.\,\eqref{eq:condition_p} can be written as
\begin{equation}
    0 \leq \tilde{u}_{i-\frac{1}{2}} - u_{p(i)-\frac{1}{2}}  < \Delta u\,. 
\end{equation}
In other words, $p(i)$ corresponds to the index of the closest neighbor of the original mesh that is smaller than $\tilde{u}_{i-\frac{1}{2}}$. Because $u$ is uniformly spaced, this calculation is straightforward
\begin{equation}
 p(i) = {\rm int} \left( \frac{ \tilde{u}_{i} - u_{0}}{{\Delta u}}\right)\,,
 \label{eq:index_u}
\end{equation}
where ``${\rm int}(a)$" represents the smaller integer that is closer to $a$. In what follows we introduce a method to build a mesh such that finding this transformation is trivial.

Consider the mesh density function $\psi:[x_{\rm min},x_{\rm max}]\to \mathbb{R}_{> 0}$ such that
\begin{equation}
\label{eq:norm_mesh}
    \int_{x_{\rm min}}^{x_{\rm max}} \psi(s) \, ds = 1\,.
\end{equation}
We define $F:[x_{\rm min},x_{\rm min}]\to [0,N-1]$ such that
\begin{equation}
F(x) = (N-1) \int_{x_{\rm min}}^{x} \psi (s)\,ds\,.
\end{equation}
The cumulative function $F$ serves as the ``index function" as it returns the (continuous) index of a mesh distributed according to the mesh density $\psi$. This correspondence is evident from the fact that the number of cells per unit space is proportional to $dF/dx = \psi$. The mesh nodes are defined implicitly by
\begin{equation}
 F( x_{i+\frac{1}{2}}) - F( x_{i-\frac{1}{2}}) =  (N-1) \int_{x_{i-\frac{1}{2}}}^{x_{i+\frac{1}{2}}} \psi (s)\,ds\, = 1\,,
\label{eq:mesh}
\end{equation}
with $x_{-\frac{1}{2}} \equiv x_{\rm min}$. 
Finding analytical expressions for $F$ or its inverse may not be possible, so numerical integration, combined with root finding methods, has to be used.
Finally, the coordinate transformation $u:[x_{\rm min},x_{\rm max}]\to [0,1]$ such that
\begin{equation}
    u(x) = \int_{x_{\rm min}}^x \psi(s)\,ds\,,
    \label{eq:u}
\end{equation}
ensures that, by definition, $\Delta u$ is constant, and Eq.\,\eqref{eq:index_u} evaluates to
\begin{equation}
 p(i) = {\rm int} [ F(\tilde{u}_i)]\,,
\end{equation}
which allows us to speed up index calculations needed in step 2 of {\method} significantly. It is worth noticing that \cite{HUGGER1993} generalizes a similar method to multiple dimensions.

\subsection{Step 3 - Obtaining $Q_{i}^{n+1}$ from $M_{i}^{n+1}$}
\label{sec:step3}
Once $M_i^{n+1}$ is obtained, the third step consists of using Eq.\,\eqref{eq:Qmass_i} to obtain the advected profile $Q^{n+1}_i$ as
\begin{equation}
Q_i^{n+1} = \frac{M^{n+1}_i}{\Delta x_i}.
\label{eq:step3}
\end{equation}

\subsection{Example of advection with RAM}
In Figure \ref{fig:method} we advect to the right a sinusoidal profile, $Q(x,t_n)$, sampled on a discrete non-uniform grid over a length $v_0\Delta t$. To advect it we follow the three steps described previously. The first step (upper panel) consists in calculating $q_i^n$, which, for simplicity, is done using PLM. These linear functions are plotted with red lines. 
To explain the process in more detail, we now focus on the interval $[x_{4-\frac{1}{2}},x_{4+\frac{1}{2}}]$ only, but a similar procedure has to be done for all the intervals. The goal is to find $Q_{4}^{n+1}$ at $x_4$ and time $t_{n+1}$ that results from advecting $q^n$. 
For that we follow step 2 (middle panel). We transform $x_4$ to $\tilde{x}_4$ and find all $p(i)$ that are within $[\tilde{x}_{4-\frac{1}{2}},\tilde{x}_{4+\frac{1}{2}}]$. This step can be done efficiently if the mesh follows a mesh-density function $\psi$ as explained in Section \ref{sec:mesh-density}.
For this particular example, the specific indices needed to evaluate Eq.\,\eqref{eq:mass_shift_algo} in this interval are $p(4),p(5)$. To satisfy the inequality \eqref{eq:condition_p}, $p(4)=1$ and $p(5)=3$, and Eq.\,\eqref{eq:mass_shift_algo} becomes
\begin{equation}
 M^{n+1}_4 = 
       \int^{x_{1+\frac{1}{2}}}_{\tilde{x}_{4-\frac{1}{2}}} q^n_{1} \,{\rm d}x + M^{n}_2   + \int^{\tilde{x}_{4+\frac{1}{2}}}_{x_{3-\frac{1}{2}}} q^n_{3} \,{\rm d}x\,.
\end{equation}
Thus, to calculate $M_4^{n+1}$, we have to integrate $q_1^n$, $q_2^n$ and, $q_3^n$ in the interval $[\tilde{x}_{4-\frac{1}{2}},\tilde{x}_{4+\frac{1}{2}}]$. Here, the integral of $q_2^n$ is replaced by $M_2^n$ simply because the step of building $q_2$ is conservative.
Finally, in step 3 (lower panel), we take $M_4^{n+1}$ and use Eq.\,\eqref{eq:step3} to calculate $Q_4^{n+1}$ (depicted with an open triangle) at $x_4$ as $Q_4^{n+1} = M_4^{n+1}/\Delta x_4$.

\subsection{General Applicability of {\method}}
{\method} can be implemented in two- and three-dimensional hydrodynamics solvers through dimensional splitting \citep[][Section 3.4]{Stone1992,BenitezLlambay2016}, and used to advect along the direction for which the bulk velocity is larger. In this case, an additional limitation to the CFL condition due to relative shear needs to be implemented \citep[see][Section 3.5.2]{BenitezLlambay2016}.
In Section\,\ref{sec:test_planet}, we show an example of a two-dimensional polar disk in which {\method} was implemented through dimensional splitting, using it to solve advection along the azimuthal direction, for which the Keplerian speed dominates the dynamics.
{\method} is also agnostic to the discretization scheme. It can be used for orbital advection in both finite difference and finite volume methods, as done previously with the uniform-mesh FARGO method \citep{Johnson2008, StoneGardiner2010,Mignone2012A&A,Lyra2017}.
We finally note that if higher order time-integration is needed on multi-dimensional problems, an extra source term needs to be considered in the hydrodynamics equations \citep[see][Eq. 10]{Shariff2018}.

\section{Testing {\method}}
\label{sec:tests}

\begin{figure*}[t]
	\centering
	 \includegraphics[scale=1.0]{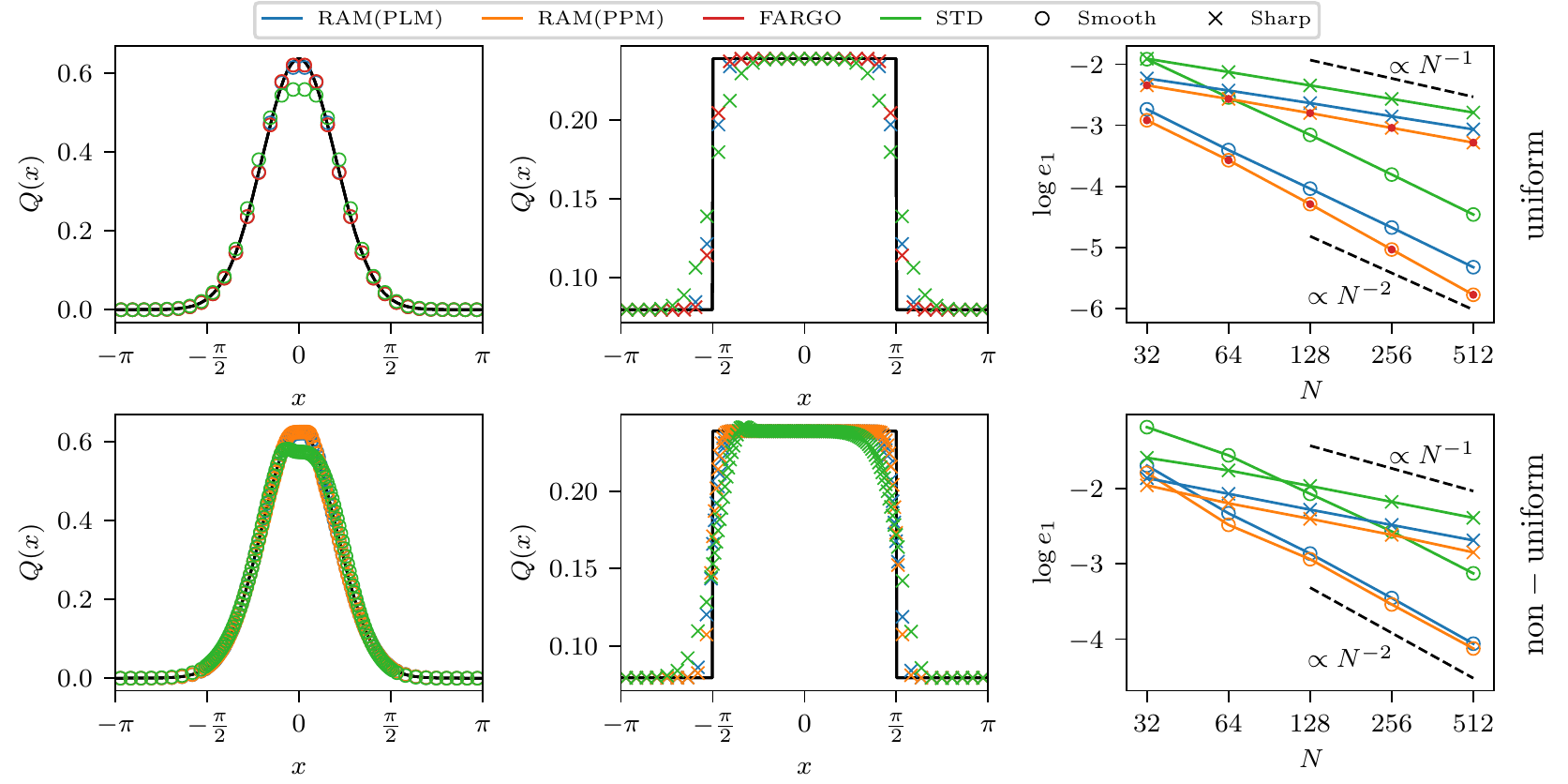}
	\caption{Results of the 1D test with RAM on uniform (upper panels) and non-uniform (lower panels) meshes described in Section \ref{sec:1d-advection}. Left and middle panels show the initial condition (black curve) and a simulation snapshot at $t=2$ (symbols). The left panel advects a smooth profile on a mesh with 32 grid points. The middle panel advects a square profile on a mesh with 128 grid points. Right panels show the errors, which for smooth profiles converge as $N^{-2}$ with both PLM and PPM reconstructions. }
	\label{fig:1d-advection}
\end{figure*}

In this section, we present results from one and two dimensional test problems, solved with the code FARGO3D \citep[][]{BenitezLlambay2016}. To test {\method} on non-uniform meshes, we extended the code to allow for non-uniform spacing along the $x$ direction in Cartesian coordinates, and $\phi$ direction in polar coordinates. Additional modifications have been added to ensure mass and momentum conservation on non-uniform meshes. We describe these modifications in Appendix \ref{ap:conservation}.
Our goal is twofold; first, we aim to characterize the convergence rate of {\method} for the interpolation methods described in Section \ref{sec:step1}. Secondly, we want to compare the efficiency of {\method} with respect to other advection methods.
These methods are the so-called standard\footnote{Enabled in FARGO3D through the option \texttt{-DSTANDARD} in the \texttt{.opt} configuration file.}(STD) and FARGO methods, the latter being the default advection algorithm used along $x / \phi$ in FARGO3D. STD solves advection through upwind methods using the full velocity \citep[see][for details]{BenitezLlambay2016}, whereas FARGO and {\method} solve advection following the splitting described in Section \ref{sec:introduction}. For both FARGO and {\method}, additional STD advection steps are required if $\delta v$ is not zero. Also, in the case of FARGO, even if $\delta v=0$, an STD step is required when $\Delta t$ is  chosen arbitrarily (as we do in Section \ref{sec:1d-advection}). 
All the methods utilize the reconstructions discussed in Section \ref{sec:step1}, where we specify which one is being used for each particular test problem (PLM or PPM).

\subsection{One dimensional advection with constant speed}
\label{sec:1d-advection}

To compare the errors and convergence rates for the different advection methods discussed above, we study the one dimensional advection of smooth and sharp profiles on uniform and non-uniform meshes. In this test, we consider the transport step of FARGO3D only \citep[see][for details]{BenitezLlambay2016}. This test highlights the errors associated with advection that a general hydrodynamics integration has for different advection algorithms.
The periodic domain spans the range $[-\pi,\pi]$. For the smooth profile, the initial condition is 
\begin{equation}
    Q(x) = \frac{2}{\pi}e^{-4x^2/\pi}\,,
    \label{eq:gaussian}
\end{equation}
whereas for the square profile is:
\begin{equation}
Q(x) = \frac{1}{\pi}
    \begin{cases}
       0.75 \quad |x| \leq \pi/2 \\
       0.25 \quad {\rm otherwise}
    \end{cases}\,.
\end{equation}
While Eq.\,\eqref{eq:gaussian} is not strictly periodic within the domain, the discontinuity produced at the edges in the initial condition is very small and does not introduce any significant error. In the two cases, the advection velocity is $v = \pi$, and it is used as the bulk velocity $v_0$ for both FARGO and {\method}. We integrate the system  for a time $t=2$, so the profile moves a full period, returning to its initial position. To test convergence with resolution, we double the number of grid points from $32$ up to $512$ and estimate truncation errors, $e_1$, through the L1-norm,
\begin{equation}
e_1 =  \frac{ \sum^{N-1}_{i=0}|Q_i - Q_i^{\rm ref}| \Delta x_i }{\sum^{N-1}_{i=0} \Delta x_i}\,,
\end{equation}
where $Q^{\rm ref}$ is a reference solution, that we take equal to the initial condition. This is because after one period, the exact solution corresponds to the same initial profile (see Eq.\,\ref{eq:advection_analytic}). We consider two cases, one in which the mesh is uniformly distributed and another where the mesh is distributed according to the periodic bump mesh density function described in Appendix \ref{sec:periodic_bump}, with parameters $a=\pi/2$, $b-a=0.1$, $c=5$ (see also Section \ref{sec:mesh-density}). This density function enables a smooth transition between two refinement levels, each with uniform resolution.
The time step for the integration is the standard CFL condition
\begin{equation}
    \Delta t = \beta \, {\rm min} \left(\frac{\Delta x_i}{v_0}\right)\,,
\end{equation}
with $\beta$ the CFL factor that is usually smaller than one because of stability constraints. We arbitrarily choose $\beta=0.5$ for STD
\footnote{It is worth noting that for this test, $\beta$ could be 1 and none of the methods would introduce truncation errors on uniform meshes. This is because the signal would be advected exactly one cell per time step and the conservative reconstruction used for all the methods (including STD) results in a shift that is exact to machine precision. 
Also, for this particular time step, the solution is not sensitive to the reconstruction method used nor its order. 
Therefore, one could also use the STD method combined with operator splitting (as done by FARGO) to advance the solution in a sub-cycle of time steps $\Delta x/v_0$, and, at the end of the cycle, advance the solution using the remainder, which would result in a significant speed up and reduction of truncation errors.},
and $\beta=5.5$ for FARGO and {\method}, which is of the order of the upper bound for the typical speed up obtained with FARGO in protoplanetary disk simulations.

In Figure \ref{fig:1d-advection} we summarize the results obtained for this test. We find that
when advecting a smooth profile, truncation errors for all the methods are second-order as they decrease quadratically with $N$ ($e_1\propto N^{-2})$, whereas for the sharp profile, the errors scale $\propto N^{-3/4}$, consistent with a first-order algorithm. For both uniform and non-uniform meshes, the absolute errors of {\method} with PLM are slightly larger than errors with PPM, as expected.
For the uniform mesh, errors using FARGO or {\method} with PPM are identical, and decrease with increasing resolution slightly faster than the errors using STD or {\method} with PLM; however the difference in slope is very small. 
The errors with {\method} (PPM) are the truncation error of the PPM interpolation. On the other hand, the errors using FARGO correspond to an additional advection with the residual velocity $v^{\rm R}_{\rm shit} = v_0 - N\Delta x/\Delta t$ (because the time step does not allow the profile to be shifted an integer number of cells only). 
This residual step is solved with the STD advection scheme with PPM reconstruction \citep[see equation (69) in][]{BenitezLlambay2016}. Thus, it is not surprising that FARGO and {\method} with PPM produce the same errors. 

Finally, the large difference between the absolute errors of STD and
 {\method}/FARGO is due to the significant difference between the time steps chosen for each method. As discussed in Section \ref{sec:introduction}, truncation errors scale with the time step and the bulk velocity $v_0$. This highlights the importance of having a method that allows advection using larger time steps on both uniform and non-uniform domains.

This first test demonstrates the validity of our formulation of {\method}. As expected, the precision and convergence properties of {\method} are controlled by the interpolation step. It also shows that {\method} is equivalent to FARGO on uniform meshes, while being its generalization on non-uniform grids.

\subsection{Planet embedded in a two-dimensional polar disk}
\label{sec:test_planet}
\begin{figure}[]
\centering
\includegraphics{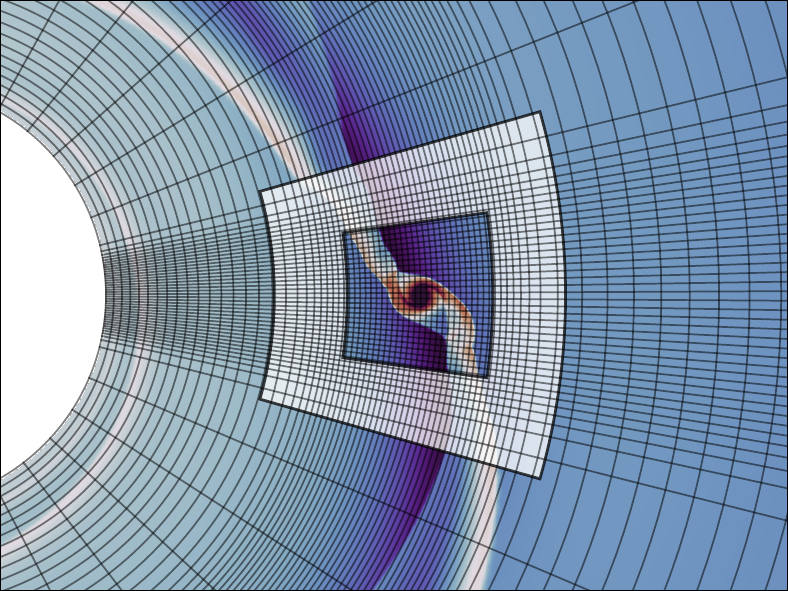}
\caption{A schematic of the polar mesh utilized in the test described in Section\,\ref{sec:test_planet}. The mesh is built in $(\phi, r)$ coordinates using the periodic bump mesh density function described in Appendix \ref{sec:periodic_bump} and Appendix\,\ref{ap:radial_bump}, respectively. The white region corresponds to the resolution transition controlled by the difference $b-a$ parameters.}
\label{fig:mesh_cyl}
\end{figure}
In this section, we demonstrate that {\method}, combined with the mesh density function, can be used to accelerate hydrodynamics calculations of differentially rotating fluids and decrease numerical diffusion on non-uniform grids.
With this purpose in mind, we set up a two-dimensional viscous disk orbiting a central star of mass $M_{\star} = M_0$. The disk model\footnote{The disk model is a slightly modified version of the publicly available setup \texttt{fargo}.} corresponds to a steady-state $\alpha$-disk \citep{Shakura73}, in which the gas surface density, $\Sigma$, and temperature, $T$, are power laws in radius with exponents -1/2 and -1, respectively.
The disk aspect ratio and viscosity parameters are set to $h = 0.05$ and $\alpha=10^{-3}$.
We use polar coordinates, where the disk domain extends radially and azimuthally as $r/r_0 \in \left[ 0.4, 2.1 \right]$, $\varphi \in \left[-\pi, \pi \right]$, respectively. 
The disk is assumed to be locally isothermal, with a sound speed $c_{\rm s} = h v_{\rm K}$, where $v_{\rm K}=\sqrt{GM_\star/r}$ is the Keplerian speed and $G$ is the gravitational constant. 
We include an embedded planet of mass $m_{\rm p} = 10^{-3} M_\star$ orbiting the central star on a circular orbit of radius $r_{\rm p}/r_0 = 1$, and modeled as a Plummer potential with small softening length $3\times 10^{-3} r_0$. 
The planet is inserted at $t=0$ and the calculations are stopped after one orbital period of the planet, at $t=2\pi \Omega_0^{-1}$, with $\Omega_0=\sqrt{GM_0/r_0^3}$ the Keplerian frequency at $r_0$.
The hydrodynamics equations are solved in a frame that co-rotates with the planet, where the planet is fixed at an azimuth $\varphi_{\rm p} = 0$. We choose a unit system such that $G = r_0 = M_0 = 1$.
The intent of this test is not to simulate planet-disk interaction realistically, but to demonstrate the efficiency of {\method} for an astrophysical scenario with high Mach number bulk velocities, high amplitude, and small-scale flow features. We  compare the computational cost of {\method} with traditional advection schemes.

We run simulations with $45$, $90$ and $180$ cells within the Hill radius $R_{\rm H} = r_{\rm p} \sqrt[3]{Gm_{\rm p}/(3 M_\star)}$ on both azimuthally uniform and non-uniform meshes. To conserve resolution per scale height across the disk domain, meshes that are evenly spaced in azimuth are also radially uniform in logarithmic space, which is a common choice for the study of planet-disk interaction. 
Runs with non-uniform spacing are distributed azimuthally according to the periodic bump mesh density function described in Appendix \ref{sec:periodic_bump}, with parameters, $a=0.1387$, $b=0.2773$ and $c=15$. 
In the radial direction, we use the modified bump of Appendix \ref{ap:radial_bump}, centered at $s_0=1$, with the same $a$ and $b$ parameters, but with $c=2.5$.
In Figure \ref{fig:mesh_cyl}, we depict the structured non-uniform mesh generated with these mesh-density functions.
The white region corresponds to the resolution transition controlled by the difference $b-a$. This mesh increases resolution close to the planet in the radial and azimuthal directions. Also, because of the logarithmic nature of the radial mesh, the cell size decreases inwards.
When the azimuthal spacing is uniform, we utilize both FARGO and STD as advection methods (we already showed in Section \ref{sec:1d-advection} that {\method} is equivalent to FARGO on uniform domains). Otherwise, when the azimuthal spacing is non-uniform, advection is solved with STD and {\method}. In all cases, we use the faster PLM reconstruction method. As demonstrated in Section \ref{sec:1d-advection}, results with PPM are similar but with lower truncation errors.
Due to the small azimuthal size of the cells in the innermost region of the disk and the large bulk speed, a {\method}-like method is expected to produce a large speed up.

\subsubsection{Theoretical speed up of {\method}}
\label{sec:speedup}

\begin{figure}[t]
\centering
\includegraphics[]{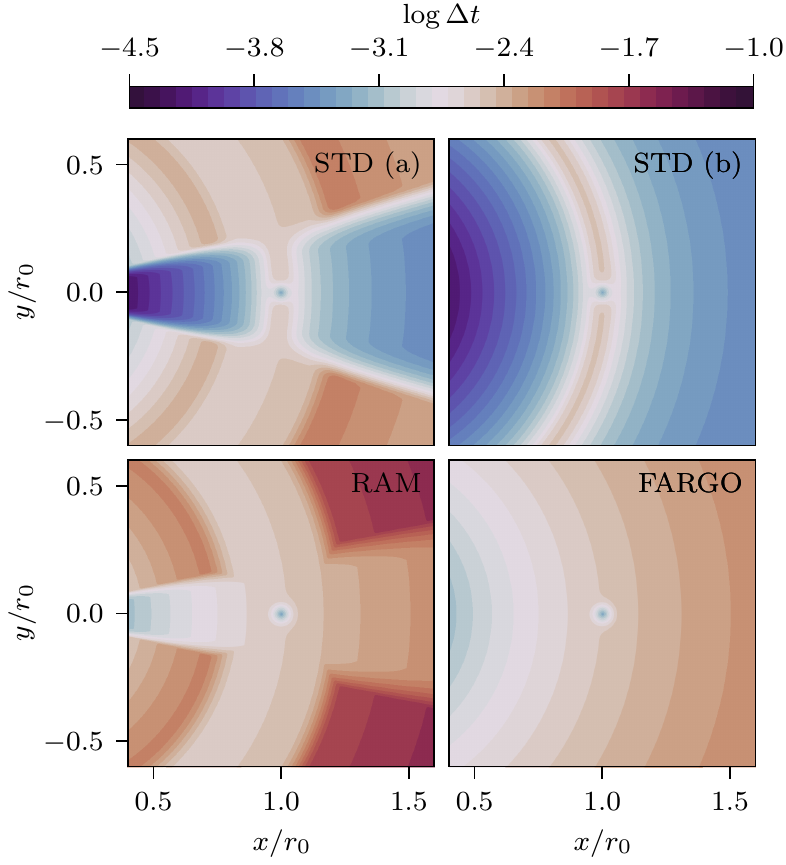}
\caption{Time step obtained with Eq.\,\eqref{eq:cfl} when applied to non-uniform meshes (left panels) and uniform meshes (right panels) with a  resolution of 180 cells/$R_{\rm H}$ at the planet location $(x,y)=(1,0)$. The upper panels show the allowed time step for STD and the lower panels for RAM or FARGO.}
\label{fig:cfl}
\end{figure}

We analyze the main terms in the CFL stability condition to obtain a first estimate of the expected computational gain using  {\method} for this particular problem. The time step is computed following Section 3.5 in
\citet{BenitezLlambay2016}. We also account for the strong gravitational acceleration produced by the planet by assuming that rotational balance can be reached quickly after the planet is inserted in the disk. In this way, $\Delta t$ may be limited by the additional local Keplerian speed induced within the Hill radius. This estimation corresponds to a lower bound for $\Delta t$ because, in reality, rotational equilibrium may take longer to be established fully or pressure support can make the disk rotate slower. The expression we use to calculate $\Delta t$ is
\begin{equation}
\begin{aligned}
\Delta t = \beta \min &\left[ \left( c_{\rm s} + v_{\rm p}\right)\max(\xi, \eta) + \xi |v_{\rm K}-v_{\rm frame}| \right.\\
&\left. + 4\nu \max(\xi^2, \eta^2) \right]^{-1}\,,
\end{aligned}
\label{eq:cfl}
\end{equation}
where $\xi = 1/(r \Delta \varphi), \eta = 1/\Delta r$, and $v_{\rm frame} = \Omega_{\rm K}(r_{\rm p}) r$ is the azimuthal velocity of the rotating frame. Following the FARGO3D implementation, we define the CFL parameter $\beta=0.44$. The rotational velocity induced by the planet, $v_{\rm p}$, is given by
\begin{equation}
v_{\rm p} = \sqrt{\frac{G m_{\rm p}}{s}} e^{-\frac{s^2}{2 \sigma^2}}\,,
\end{equation}
with $s = |{\bf r}-{\bf r}_{\rm p}|$ and $\sigma = R_{\rm H}/2$. Here, the Keplerian speed has been multiplied by an exponential cutoff to make the effect of rotation local. 

Eq.\,\eqref{eq:cfl} includes the contribution of the bulk velocity of the disk, $v_{\rm K}-v_{\rm frame}$, and the velocity induced by the planet, $v_{\rm p}$. As such, for a given mesh, it estimates $\Delta t$ for STD. To obtain the time step allowed by {\method} or FARGO, we set $v_{\rm K}-v_{\rm frame}=0$. In addition, the maximum speed up possible by RAM can be calculated after setting $v_{\rm p} = 0$.
In Figure \ref{fig:cfl}, we show the results of Eq.\,\eqref{eq:cfl} when applied to non-uniform (left panels) and uniform (right panels) meshes with 180 cells/$R_{\rm H}$. For each mesh, upper and lower panels show the allowed time step in a region that includes both the planet and the inner boundary for STD and {\method} or FARGO, respectively. 
Since $c_{\rm s}$ increases inwards while the cell size decreases (c.f., Figure \ref{fig:mesh_cyl}), $\Delta t$ decreases towards the inner boundary. Also, because in the non-uniform mesh there is an azimuthal band in which the azimuthal size also decreases with decreasing $r$, $\Delta t$ can become even more restrictive. This effect is particularly strong for STD which also has the contribution of the bulk velocity, $v_{\rm K}-v_{\rm frame}$, that increases rapidly for decreasing $r$. 
Another remarkable feature is the small $\Delta t$ produced due to the high rotational velocity induced by the planet within a high-resolution region around $(x,y)=(r_0,0)$. 
Whether or not this velocity dominates the time step of the simulation depends exclusively on the location of the inner boundary (the closer it is to the central star, the smaller the time step).
When using {\method}, the absence of the term proportional to the bulk velocity allows $\Delta t$ to be larger almost everywhere, except at the planet's orbit, where the bulk velocity is, by construction, zero.
Therefore, in the case of {\method}, the terms that limit $\Delta t$ are those associated with $c_{\rm s}$ and $v_{\rm p}$. The former may limit $\Delta t$ due to the double effect of $c_s$ being large inwards and the cell size being also small there, while the latter may limit $\Delta t$ due to the small cell size and large velocity induced close to the planet.
In the case in which the mesh is uniform, we also calculate $\Delta t$ for FARGO and STD. The time step allowed globally by this mesh is almost indistinguishable from the one obtained for the non-uniform case. The same dependencies of $\Delta t$ with $r$ are observed.
It is also worth noticing that, for both meshes, viscosity could limit $\Delta t$ strongly if either $\alpha$ or the resolution is further increased.

We now report the maximum allowed $\Delta t$ obtained for {\method} and STD with/without a planet on the non-uniform mesh. For STD with planet, the maximum time step is $\Delta t_{\rm STD} = 5.7\times10^{-5} \Omega_0^{-1}$ while for {\method} we obtain $\Delta t_{\rm RAM} = 2.9\times10^{-4} \Omega_0^{-1}$. This shows that {\method} allows a time step that is $\sim5$ times larger than STD. If we do not consider the planet in this calculation, for STD we obtain $\Delta t_{\rm STD, np} = 5.7\times10^{-5} \Omega_0^{-1}$ and $\Delta t_{\rm RAM, np} = 7.6 \times 10^{-4} \Omega_0^{-1}$ for {\method}, which is $\sim13$ times larger than that of STD.

Finally, when considering a uniform mesh, producing the same resolution of 180 cells/$R_{\rm H}$ requires sixteen times more cells than in the non-uniform case, using eight and two times more cells in azimuth and radius, respectively. Given that {\method} and FARGO produce the same time step, {\method} on a non-uniform mesh is sixteen times faster than a FARGO run on a uniform mesh.
When comparing {\method} on non-uniform meshes with respect to STD on a uniform one, {\method} can be $\sim80-208$ times faster than STD on a co-rotating frame, and even faster in non-corotating frames.

\subsubsection{Simulation results}
\label{sec:simulations}

\begin{figure}[t]
\centering
\includegraphics[]{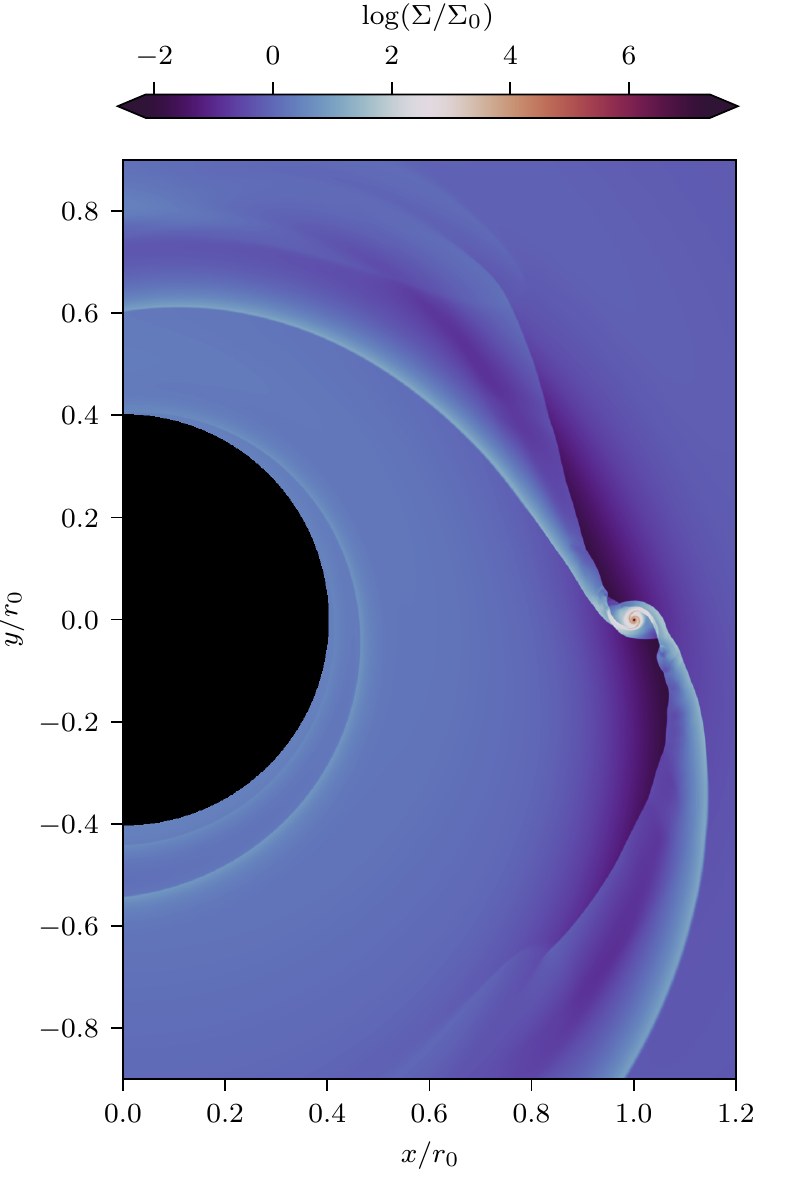}
\caption{Snapshot of the 2D global run for the RAM method with 180 cells/$R_{\rm H}$. The non-uniform mesh enables to resolve the small-scale perturbation near the planet as well as the spiral wakes propagating in the disk. A zoomed version of the plot is shown in Figure \ref{fig:planet}. }
\label{fig:planet_global}
\end{figure}

\begin{figure*}[t]
\centering
\includegraphics[]{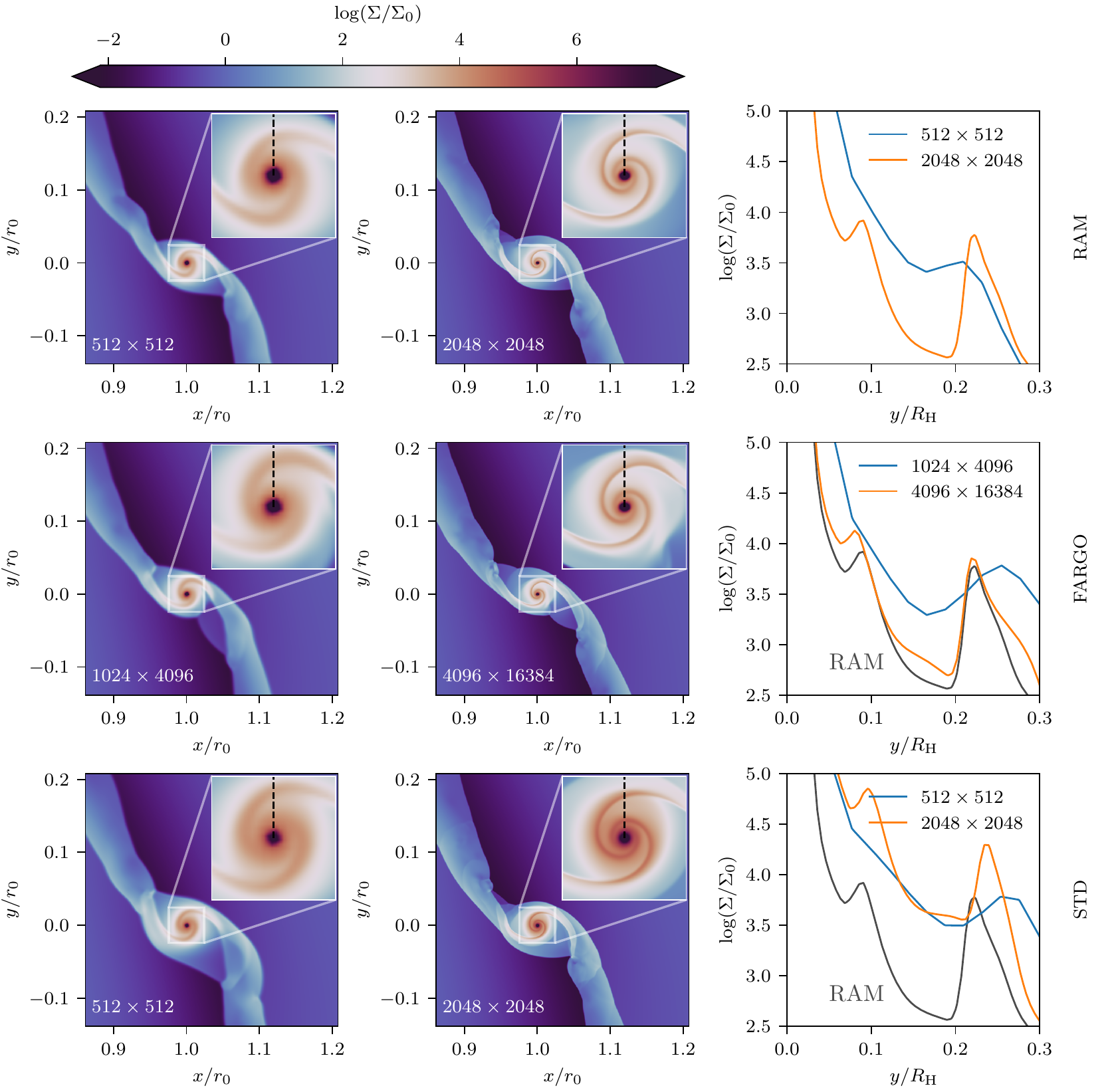}
\caption{ Snapshots of surface density after one orbit for runs with 45 cells/$R_{\rm H}$ (left panels) and 180 cells/$R_{\rm H}$ (middle panels), for {\method} (upper panels),
FARGO with a uniform mesh (middle panels), and STD in a non-uniform domain (lower panels), where we also include, for each resolution, one zoom-in panel within the circumplanetary region (the global scale of the runs is shown in Figure \ref{fig:planet_global}). Rightmost panels show a cut of surface density along the solid line drawn in the zoom-in for each resolution (blue/orange curve corresponds to the low/high-resolution run). We plot the high-resolution RAM run with solid gray lines to facilitate the comparison between the methods. STD is clearly affected by larger numerical diffusion, whose main effect is to smooth out the spirals and to increase the ambient density within the circumplanetary region. Given that the bulk velocity is zero at the planet's orbit, the extra diffusion for STD is only explained by the smaller time step used and the corresponding larger number of hydrodynamics steps (which translates into a larger number of interpolations/reconstructions before reaching the output time). When comparing the circumplanetary region of {\method} and FARGO, we observe a remarkable agreement between the methods, but with {\method} being $\gtrsim 10$ times faster.}
\label{fig:planet}
\end{figure*}

On large scales, the features obtained with different methods/resolutions are indistinguishable from one another. We highlight these features in Figure \ref{fig:planet_global}, that shows the surface density, $\Sigma$, for the run advected with {\method} and 180 cells/$R_{\rm H}$.
In Figure \ref{fig:planet}, we show snapshots $\Sigma$ for the runs with 45 cells/$R_{\rm H}$ (left column) and 180 cells/$R_{\rm H}$ (middle column), for {\method} (upper panels), FARGO (middle panels), and STD in a non-uniform domain (lower panels) after one orbit of the planet.
In each panel, we zoom-in the circumplanetary region where spirals develop. Also, in the rightmost panels we plot cuts along the black line drawn in the zoom-in for different resolutions (low/high resolution with blue/orange curves).

It is when we look within the circumplanetary region that significant differences are observed for runs with low and high resolution. The ambient density within the Hill radius and the density contrast of the spirals are most impacted. Low-resolution runs are affected by larger truncation errors and numerical diffusion.
The effect of these errors is to increase the ambient density within $R_{\rm H}$ and smooth out the spirals. This can be recognized when comparing low-resolution runs with FARGO, {\method} and STD with both, low and high-resolution (blue curves in the rightmost panels). 
We note that the high numerical diffusion of STD within the circumplanetary region is not due to the bulk velocity because this velocity is small in the corotating frame used. So, we conclude that the smaller time steps of STD and the corresponding larger number of interpolations/reconstructions required to advance the solution up to the final integration time are the main source of numerical diffusion close to the planet.
It is worth noting that {\method} on a non-uniform mesh reproduces the sharp features obtained with FARGO. 
However, the non-uniform mesh utilizes sixteen times fewer grid points than FARGO, 
significantly reducing the computational cost while maintaining the same level of accuracy.

We now report measurements of the time step required for each method in different meshes for the high-resolution run with 180 cells/$R_{\rm H}$ only, but similar results were obtained for the other runs with lower resolution.
We find that $\Delta t_{\rm STD} = 5.7 \times 10^{-5} \Omega^{-1}_0$ and $\Delta t_{\rm RAM} = 3.87 \times 10^{-4} \Omega^{-1}_0$. 
Despite the simplified modeling, these values are in excellent agreement with the estimations described in Section\,\ref{sec:speedup}.
Ideally, the relative speed up between methods is given by the ratio $\Delta t_{\rm RAM}/\Delta t_{\rm STD} \simeq 6.9$, however {\method} introduces an overhead with respect to STD due to the velocity splitting (see Section \ref{sec:introduction}) and the additional calculations required to shift the profile (see Section \ref{sec:method}).
Using fast index calculation (see Section \ref{sec:mesh-density}), in our tests we found that this overhead is $\sim 20\%$. Therefore, the effective speed up is  $\Delta t_{\rm RAM}/\Delta t_{\rm STD} \times 0.8 \simeq 5.5$.
Note that a similar overhead is introduced with FARGO, but it requires eight times more cells in azimuth to match the resolution of 180 cells/$R_{\rm H}$. Since we find $\Delta t_{\rm RAM} \simeq \Delta t_{\rm FARGO}$, we conclude that FARGO is eight times more expensive than {\method} on non-uniform meshes provided the radial spacing used for both methods is identical. However, in uniform meshes like the one presented in Section\,\ref{sec:test_planet}, where the number of cells is two and eight times more in radius and azimuth, respectively, FARGO is sixteen times more expensive than {\method} on the non-uniform mesh.

The problem studied in this section illustrates the main advantages of {\method} to study the dynamics close to a planet at a lower computational cost. Two main results can be highlighted from our analysis. First, the standard approach using FARGO to simulate the problem on an azimuthally uniform mesh is expensive. Interestingly, a cheaper method consists in utilizing STD on a non-uniform mesh (both in radius and azimuth). Second, our results clearly show the advantage of using non-uniform meshes combined with {\method}, which has smaller truncation errors and is significantly cheaper than STD.

\section{Summary and Conclusions}
\label{sec:conclusions}

In this work, we developed {\method} to solve the advection equation that allows larger time steps and lower truncation errors with respect to standard up-wind explicit advection algorithms in problems dominated by a large bulk velocity.
Our method is based on operator splitting and conservative interpolation with monotonic reconstruction. 
We provided PLM \citep{VANLEER1974} and PPM \citep{Colella1984} formulas to implement {\method} obtaining a method that can be first or second-order accurate, but other reconstruction methods could be used.
{\method} works with arbitrary meshes, but we provide a recipe to generate meshes through mesh-density functions that allow faster interpolation.
It can be applied to multi-dimensional problems via dimensional splitting. In this case, an additional time step limitation arising from local shear needs to be implemented \cite[see][]{Masset2000, BenitezLlambay2016}. Higher order (in time) implementations of {\method} require an additional source term \citep[see][]{Shariff2018}.

We tested {\method} by implementing it on the publicly available code FARGO3D. By advecting sharp and smooth one-dimensional profiles, we measured the convergence 
rate of our method for different reconstructions and compared them with FARGO and standard up-wind advection on uniform and non-uniform meshes. Smooth profiles converged  faster than sharp profiles for all the methods. Also, this test demonstrated that FARGO and {\method} give identical results on uniform meshes. On non-uniform meshes, {\method} converges towards the right solution while being cheaper and more precise than standard advection.

Motivated by the long-standing problem of planet-disk interaction, we also tested {\method} in a two-dimensional polar disk with very high resolution within the Hill radius of an embedded planet.
For this problem, we found {\method} is $\sim 5.5$ times faster than standard advection on non-uniform meshes. When comparing {\method} on a non-uniform mesh with respect to FARGO on a uniform one with the same effective resolution within the Hill radius, we found that {\method} required  $1/16^{th}$ the computational cost of FARGO.
We emphasize that the speed up or computational cost reduction obtained with {\method} is problem-dependent. Since {\method} and FARGO allow the same time step (for the same effective resolution), the reduction in computational resources required scales with the ratio between the number of grid points used (i.e., computational overhead).

While in this paper we have worked with meshes generated by a mesh-density function, {\method} can be applied to arbitrary meshes, including  nested meshes, or under the framework of adaptive mesh refinement. The utility of one mesh-generating method over another is surely problem dependent.

We conclude that {\method} reduces the computational cost of hydrodynamics calculations significantly while decreasing truncation errors in problems dominated by a large bulk velocity. This, combined with the possibility of applying the method to non-uniform meshes, allows the numerical cost of hydrodynamics to be decreased by more than one order of magnitude. The actual value depends on the bulk velocity.
This method may impact several fields concerning studies of the so-called disk-satellite problem. In particular, it can be used to speed up two- and three-dimensional calculations of planet-disk interaction in problems that require both global disks and high resolution within the Hill radius of the planet and planetary orbit.

\vspace{0.5cm}
We thank the anonymous referee whose comments helped us improve this manuscript. We thank Frédéric Masset for a thorough reading of an early version of this manuscript. P.~B.~L. acknowledges  support from ANID, QUIMAL fund ASTRO21-0039 and FONDECYT project 1231205. L.\ K.\ and K.~M.~K  acknowledge support  from the Heising-Simons 51 Pegasi b postdoctoral fellowship and TCAN grant 80NSSC19K0639. X.~S.~R. acknowledges support from ANID, Millennium Science Initiative, ICN12\_009.

\appendix
\section{PPM}
\subsection{Interface interpolation for the PPM method}
\label{ap:PPM}
Following \citet{Colella1984} $\langle Q\rangle_{i+\frac{1}{2}}$ are calculated as
\begin{align}
    \langle Q\rangle_{i+\frac{1}{2}} &= \langle Q\rangle_{i} + \frac{\Delta x_i}{\Delta x_i+\Delta x_{i+1}} \left(\langle Q\rangle_{i+1} -\langle Q\rangle_{i} \right)
     + \frac{ \bigg( \tilde{\omega}_1 (\langle Q\rangle_{i+1}-\langle Q\rangle_{i}) + \tilde{\omega}_2  \delta \langle Q\rangle_{i+1} + \tilde{\omega}_3 \delta \langle Q\rangle_{i}  \bigg) }{\sum^2_{k=-1} \Delta x_{i+k}} 
\end{align}
with
\begin{equation}
\begin{aligned}
     \tilde{\omega}_1 &= \frac{2 \Delta x_{i+1}\Delta x_i}{\Delta x_i+\Delta x_{i+1}} \left[ \frac{\Delta x_{i-1} + \Delta x_{i}}{2\Delta x_{i}+\Delta x_{i+1}} - \frac{\Delta x_{i+2}\Delta x_{i+1}}{2\Delta x_{i+1}+\Delta x_{i}} \right]  \\
     \tilde{\omega}_2 &= -\Delta x_i \frac{\Delta x_{i-1}\Delta x_{i}}{2\Delta x_{i} + \Delta x_{i+1}}  \\
     \tilde{\omega}_3 &= \Delta x_{i+1} \frac{\Delta x_{i+1}\Delta x_{i+2}}{\Delta x_{i} + 2\Delta x_{i+1}} 
\end{aligned}
\end{equation}
and
\begin{equation}
\begin{aligned}
    \delta \langle Q\rangle_{i} &= \frac{\Delta x_i}{\Delta x_{i-1}+\Delta x_{i}+\Delta x_{i+1}} \left( \frac{2\Delta x_{i-1} + \Delta x_{i}}{\Delta x_{i}+\Delta x_{i+1}} \left(\langle Q\rangle_{i+1} -\langle Q\rangle_{i}\right)
    +  \frac{\Delta x_{i} + 2\Delta x_{i+1}}{\Delta x_{i-1}+\Delta x_{i}} \left(\langle Q\rangle_{i}-\langle Q\rangle_{i-1}\right) \right)
\end{aligned}
\end{equation}

\subsection{PPM Monotonicity constraint}
\label{ap:monotonocity}

To apply the monotonicity constraint, first $\delta \langle Q\rangle_i$ is modified in the calculation $\langle Q\rangle_{i+\frac{1}{2}}$ as
\begin{equation}
    \delta \langle Q\rangle_{i} =  
    \begin{cases}
       {\rm min}\left(|\delta \langle Q\rangle_i|, 2|\langle Q\rangle_{i+1} - \langle Q\rangle_{i}|, 2|\langle Q\rangle_{i}-\langle Q\rangle_{i-1}| \right) \quad \text{if } \left(\langle Q\rangle_{i+1} - \langle  Q\rangle_{i}\right)\left(\langle Q\rangle_{i} - \langle Q\rangle_{i-1}\right) > 0 \nonumber \\
       0  \quad \text{otherwise }
\end{cases}\,.
\end{equation}
With updated values of $\langle Q\rangle_{i+\frac{1}{2}}$ the coefficients $Q_{\rm R,i}$ and $Q_{\rm L,i}$ are obtained as
\begin{align}
    Q_{\rm R,i} = Q_{\rm L,i} = \langle Q\rangle_{i}  \quad\,\, \quad \quad  &\textrm{if } \left(Q_{\rm R,i} - \langle Q \rangle_{i}\right)\left( \langle Q\rangle_{i} - Q_{\rm L,i}\right) \leq 0 \\
    Q_{\rm L,i} = 3\langle Q\rangle_{i} - 2 Q_{\rm R,i}  \,\, \quad \quad  &\textrm{if } \left( Q_{\rm R,i}- Q_{\rm L,i} \right) Q_{a,i} > \frac{1}{6}\left(  Q_{\rm R,i}- Q_{\rm L,i} \right)^2 \\
     Q_{\rm R,i} = 3\langle Q\rangle_{i} - 2 Q_{\rm L,i}  \,\, \quad \quad  &\textrm{if } -\frac{1}6{}\left(  Q_{\rm R,i}- \langle Q_{\rm L,i} \right)^2 >  \left(  Q_{\rm R,i}-  Q_{\rm L,i} \right) Q_{a,i}  
\end{align}
with $Q_{a,i} =  \langle Q \rangle_i - \frac{1}{2}\left( Q_{L,i} + Q_{R,i} \right) $.

\section{Mesh-density functions}
\label{sec:examples_mesh_density}

\subsection{Periodic bump}
\label{sec:periodic_bump}

\begin{figure*}[t]
\centering
\includegraphics[scale=0.75]{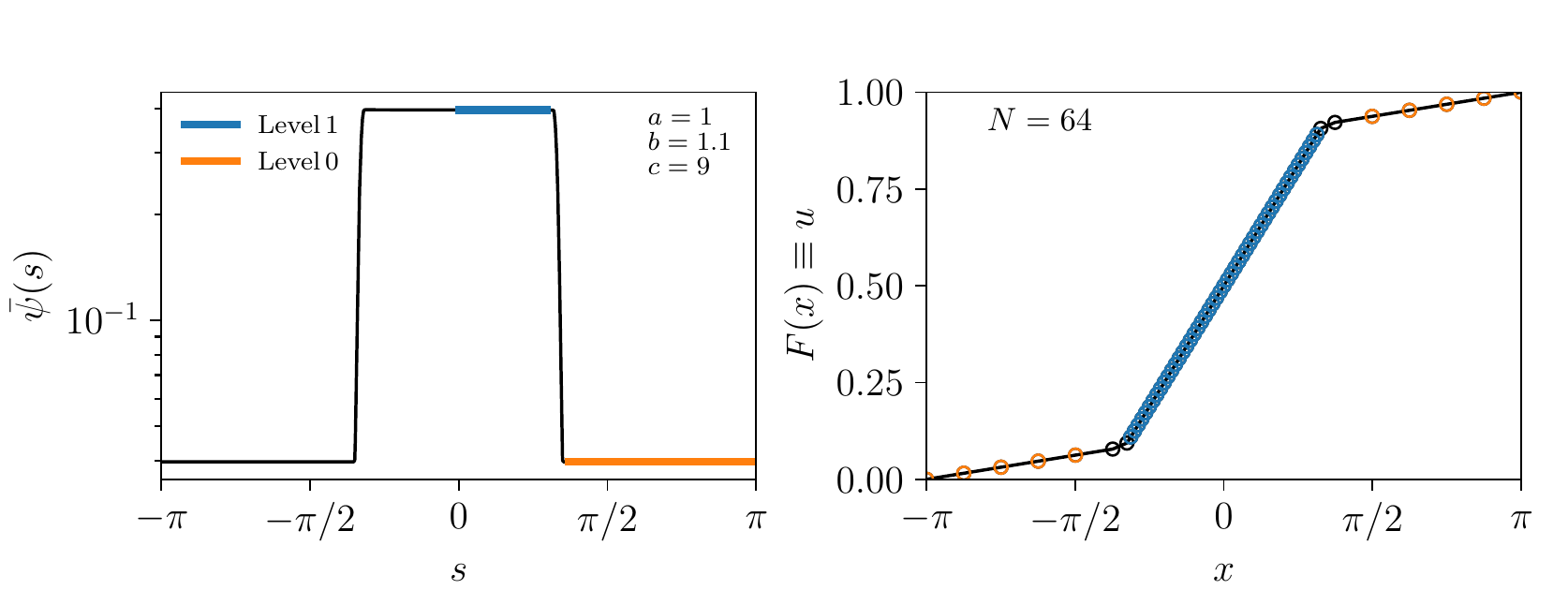}
\caption{Example of the mesh density functions.  Left panel corresponds to the periodic bump function with $a=1.0$, $b=1.1$ and $c=9$. In the right panel we plot the cumulative function, $F(x)$ as a function of the non-uniform coordinate $x$ for a mesh with $N=64$ grid points. This mesh was used in the test described in Section\,\ref{sec:1d-advection}.}
\label{fig:meshes}
\end{figure*}

In this work, we restrict our analysis to functions $F$ with known analytical expressions for the integrals and use the bisection method to solve Eq.\,\eqref{eq:mesh} or the expression of the inverse when it is easy to find.

We define the function $\psi(s)$ as
\begin{equation}
\psi(s) = \left\{
        \begin{array}{ll}
            1+c  \quad &|s| \le a \\
            1+c\cos^2\left(\displaystyle{\frac{\pi}{2} \frac{|s|-a}{b-a}}\right)  \quad a < &|s| < b \\
            1  \quad &|s| \geq b \\           
        \end{array}
    \right.
\end{equation}
with $x_{\rm min} \leq s \leq x_{\rm max}$. 
To satisfy the condition given by Eq.\,\eqref{eq:norm_mesh} we normalize $\psi(s)$ by $I_\psi$, which is obtained after integrating the density function over the domain $\left[x_{\rm max}, x_{\rm min}\right]$, that is
\begin{align}
\label{eq:Inorm}
    I_\psi = \int_{x_{\rm min}}^{x_{\rm max}} \psi(s) ds = (x_{\rm max}-x_{\rm min}) + c\left(a+b\right)\,.
\end{align}
Thus, the desired mesh density function corresponds to
\begin{equation}
\bar{\psi} = I_\psi^{-1} \psi.    
\end{equation}
This function generates a mesh with two levels of uniform resolution and includes a smooth transition between the two levels. 
The value of the parameter $a$ sets the extent of the highest resolution level. The parameter $b$ sets the location of the transition between two levels, and therefore $|b|>|a|$. 
The extent of the transition is given by $b-a$. 
The value of the parameter $c$ sets the relative resolution between the two levels. 

For this case, the cumulative $F(x) = \int_{-\pi}^x \bar{\psi}(s) ds$, which corresponds to the definition of the coordinate $u(x)$, is:
\begin{align}
F(x) &= \left\{
\begin{array}{ll}
f(x) - f(-\pi) &\quad  x \leq -b \\
g^+(x) - g^+(-b) +  F(-b)  &\quad -b < x \leq -a \\
h(x) - h(-a) + F(-a)&\quad -a < x \leq a \\
g^-(x) - g^-(a) +  F(a)  &\quad a < x \leq b \\
f(x) - f(b) +  F(b)  &\quad x > b
\end{array}
\right.
\end{align}
with the functions $f, g^{\pm}, h$ defined as 
\begin{align}
f(x) &= \displaystyle{\frac{x}{I}}\,, \\
g^{\pm}(x) &= \frac{\left(2+c\right) }{2} f(x) - \frac{c\left(b-a\right)}{2\pi I } \sin\left(\pi \frac{x \pm\ a}{a-b} \right)\,, \\
h(x) &= \left(1+c\right) f(x)\,,
\end{align}

In Figure \ref{fig:meshes} we plot an example of a mesh-density function with $s \in \left[-\pi,\pi \right]$. 
The parameters are defined to provide a $\sim 5 \times$ resolution contrast between the two levels.
In the right panel of  Figure \ref{fig:meshes} we include the cumulative distribution for a mesh with $N=64$ points.
Finally, we note that the piece-wise density function proposed here can be modified to allow for arbitrary levels of resolution with continuous transitions.

\subsection{Radial bump}
\label{ap:radial_bump}

In Section\,\ref{sec:test_planet} we utilize a mesh density function in the radial direction centered at $s_0=1$ and defined as
\begin{equation}
\psi(s) = \left\{
        \begin{array}{ll}
            1/s+c  \quad &|s-s_0| \le a \\
            1/s+c\cos^2\left(\displaystyle{\frac{\pi}{2} \frac{|s-s_0|-a}{b-a}}\right)  \quad a < &|s-s_0| < b \\
            1/s  \quad &|s-s_0| \geq b \\           
        \end{array}
    \right.
\end{equation}
with $I_{\psi} = \log(x_{\rm max}) - \log(x_{\rm min}) + c(a+b)$ (see Eq.\,\ref{eq:Inorm}).
Note that when the parameter $c=0$ we obtain a log-uniform mesh density.

\section{Momentum and mass conservation in non-uniform meshes}
\label{ap:conservation}
Here we describe modifications that need to be done to the standard algorithms described in \citep[][]{BenitezLlambay2016} to conserve mass and momentum to machine precision in non-uniform meshes.
Lets consider a 3D Cartesian mesh with coordinates $x,y,z$ and corresponding cell-center index $i,j,k$.
Momentum (along $y$-direction) direction is defined as  
\begin{align}
\Pi_{y,ijk} = \frac{1}{2}\rho_{ijk} \left(v_{y,ij-\frac{1}{2}k} + v_{y,ij+\frac{1}{2}k} \right),
\end{align}
where $v_{y,j-\frac{1}{2}}$ is the velocity along the $y$-direction at the cell interface and $\rho_j$ is the density at the cell center.  

The mass of a cell corresponds to
\begin{equation}
    M_{ijk} = \rho_{ijk} V_{ijk}\,
\end{equation}
where $V_{ijk}$ is the volume of the grid cell.
Conservation of momentum and mass implies that $\sum_j \Pi^{n+1}_{y,ijk}V_{ijk} = \sum_j \Pi^n_{y,ijk} V_{ijk}$ and  $\sum_k \sum_{j} \sum_i M^{n+1}_{ijk} = \sum_k \sum_j \sum_i M^n_{ijk}$, respectively.
Updates in density and velocity must be consistent with these conservation conditions. 
Therefore we revisit the updates performed during the source and transport defined in \cite{BenitezLlambay2016}.

\subsection{Source step}

Consider a 1D pressure gradient along the $y$-direction. The velocity is updated as
\begin{align}
v_{ij-\frac{1}{2}k}^{n+s} = v_{ij-\frac{1}{2}k}^n - \frac{ \Delta t}{\left<\rho^n\right>_{ij-\frac{1}{2}k} } \frac{P^n_{ijk} - P^n_{ij-1k}}{y_j - y_{j-1}}
\end{align}
where $n+s$ indicates time-update from step $n$ and  $\left<\rho^n\right>_{ij-\frac{1}{2}k}$ is a spatial averaged density \citep[see][]{BenitezLlambay2016}.
It can be shown that momentum is conserved in uniform and non-uniform grids if the average density is defined as
\begin{align}
\left<\rho^n\right>_{ij-\frac{1}{2}k} = \frac{1}{2} \frac{\rho^n_{ijk} \left( y_{j+\frac{1}{2}} - y_{j-\frac{1}{2}} \right) + \rho^n_{ij-1k}\left( y_{j-\frac{1}{2}} - y_{j-\frac{3}{2}} \right)}{y_j - y_{j-1}}
\end{align}
The same holds for Cylindrical and Spherical coordinates. Note that the same update needs to be done in \verb|substep2| for artificial viscosity calculations.

\subsection{Transport step}
Consistency in the transport step is crucial for the conservation of mass and momentum. In non-uniform meshes, Eq.\,(50) in \cite{BenitezLlambay2016} must be updated to
\begin{equation}
v_{y,ijk}^{n+1} = \frac{\Pi^{-}_{y,ijk}V_{ijk} +  \Pi^{+}_{y,ij-1k}V_{ij-1k}}{\rho_{ijk}V_{ijk} + \rho_{ij-1k}V_{ij-1k} }.
\end{equation}
The same must be done for $v_x, v_z$. 

\end{document}